\RequirePackage{fix-cm}
\documentclass[14pt]{article}
\usepackage{makeidx}
\usepackage{multirow}
\usepackage{multicol}
\usepackage[dvipsnames,svgnames,table]{xcolor}
\usepackage{graphicx}
\usepackage{epstopdf}
\usepackage{ulem}
\usepackage{hyperref}
\usepackage{amsmath}
\usepackage{amssymb}
\usepackage{latexsym}
\usepackage[misc]{ifsym}
\usepackage{mathptmx}
\usepackage[paperwidth=595pt,paperheight=841pt,top=72pt,right=90pt,bottom=72pt,left=90pt]{geometry}

\makeatletter
	{\par\setlength{\parindent}{#3}
	\setlength{\leftmargin}{#1}       \setlength{\rightmargin}{#2}%
	\advance\linewidth -\leftmargin       \advance\linewidth -\rightmargin%
	\advance\@totalleftmargin\leftmargin  \@setpar{{\@@par}}%
	\parshape 1\@totalleftmargin \linewidth\ignorespaces}{\par}%
\makeatother

\begin{document}

\title{A Verifiable Quantum Secret Sharing Scheme Based on a Single Qubit}

\author{Dan-Li Zhi \and Zhi-Hui Li\thanks{\Letter lizhihui@snnu.edu.cn}\and Zhao-Wei Han \and Li-Juan Liu }

\footnotetext{College of Mathematics and Information Science, Shaanxi
Normal University, Xi'an 710119,China}

\date{Received: date / Accepted: date}

\maketitle

\begin{abstract}
To detect frauds from some internal participants or external
attackers, some verifiable threshold quantum secret sharing schemes have been
proposed. In this paper, we present a new verifiable threshold structure based on
a single qubit using bivariate polynomial. First, Alice chooses an asymmetric
bivariate polynomial and sends a pair of values from this polynomial to each
participant. Then Alice and participants implement in sequence unitary
transformation on the $d$-dimensional quantum state based on unbiased bases,
where those unitary transformations are contacted by this polynomial. Finally,
security analysis shows that the proposed scheme can detect the fraud from
external and internal attacks compared with the exiting schemes and is comparable
to the recent schemes.
\paragraph{keywords}Verifiability \and  Mutually unbiased bases \and Unitary transformation \and Secret sharing scheme
\end{abstract}

\section{Introduction}
\label{intro}

Quantum secret sharing is an important issue in quantum cryptography, which
combines classical secret sharing and quantum theory and plays an important role
in applied cryptography. It means that classic or quantum secret information can
be divided into shares by a dealer among  participants, so that
only authorized participants can recover the secret, and any one or
more unauthorized participants can not recover the secret. In a $(t,n)$-threshold
quantum secret sharing scheme, the dealer splits secret into $n$ shares sending them to each of
 $n$ participants where any set of $t$ or more shareholders can recover the shared
secret cooperatively, however less than $t$ shareholders can not recover it.
Hillery et al.[1] firstly proposed quantum secret sharing (QSS) based on the
quantum correlation of Greenberger-Horne-Zeilinger (GHZ) states in 1999. The main
idea of this scheme is that an unknown quantum state is shared between two
participants and only restored collaboratively. The fundamental theory of quantum
determines that the QSS schemes are more secure than the classic ones. There are
many QSS schemes [2-9] that have been proposed. For example, a quantum secret
sharing scheme was constructed using the product state rather than the
entangled state in [4], such that the scheme is applicable when the number
of participants is expanded. In addition, a $(n,n)$-threshold quantum secret
sharing scheme that shares multiple classic information was proposed based on a
single photon in [6]. Most of these schemes do not take into account two major
security issues: during the secret distribution phase, an attacker may
impersonate the dealer to send false information to the shareholders and during
the recovery phase, some participants may provide false shares, so they cannot
recover the correct secret. In 1985, Chor et al. [10] proposed the concept of
verifiable secret sharing (VSS) and gave a complete scheme that
verifiable problem was solved effectively. The verifiable secret sharing scheme has
attracted the attention of many scholars, because it can prevent dishonest
participants from providing false information during the secret recovery stage,
and it can also prevent external fraud from pretending to send false
information to participants. With the advent of quantum algorithms [11, 12], the
theory of quantum secret sharing (VQSS) [7-9] can be further developed. For
example, an identity-based quantum signature encryption algorithm was constructed
in [8], which makes the secret share and signature are safe under the choice
of plaintext attack. What's more, a verifiable $(t,n)$-threshold QSS scheme with
sequential communication was proposed in [9], in which the property of mutually
unbiased bases is used [13]. The quantum state is measured with the basis
${\left\{ {\left| {{\varphi _l}^0} \right\rangle } \right\}_l}$ by the last
participant ${\rm{Bo}}{{\rm{b}}_t}$ until the dealer Alice and all participants
${\rm{Bo}}{{\rm{b}}_1},{\rm{Bo}}{{\rm{b}}_2}, \cdots ,{\rm{Bo}}{{\rm{b}}_t}$
implement unitary transformation on the three single-bit quantum states
sequentially. Then the measurement result is sent to each participant. In the secret
recovery phase, participants exchange the information of unitary transformation
so that each participant can recover secrets, where the first two qubits
are used to share the two secrets, and the third single qubit is used as the verification
information. This scheme combines the Shamir secret sharing scheme with
sequential communication in the $d$-dimensional quantum system, and
adds the third qubit as the authentication information, which easily identifies
the spoof or attack. However, the scheme has two drawbacks: Firstly, when
participants exchange classic information, they are vulnerable to external
attackers, which will result in the scheme to be unsafe. Secondly, the third
qubit used as verification information is wasted and the verification
formula $p_0^1 = p_0^2p_0^3\bmod d$ will be established with a certain
probability, even if there are dishonest participants.

In this paper,the scheme in Ref.[9] is improved on these two
issues and a new QSS scheme based on the property of mutually unbiased bases [13]
is proposed. Compared with the original scheme, this paper has the following
advantages:
\par{(a).During the secret distribution phase, the distributor chooses an asymmetric
binary polynomial to replace the original unary polynomial. The classical
information of the participants' unitary transformation is obtained by their
respective secret shares, which will be supervised by the dealer so as to make
sure that participants are honest. Consequently, the verifiability of the scheme
is achieved with less authentication information.}
\par{(b).This scheme is supervised by
the dealer to ensure that each participant is honest,
thereby realizing the practicality and feasibility of the scheme.}
\par{(c).If a
dishonest participant is found during the secret sharing phase, the
dishonest participant will be found and removed. In the next secret sharing
process, the dealer simply re-prepares a binary polynomial and a quantum state to
recover the secret, which will prevent internal attacks.}
\par{(d).The classic information
exchanged by the participants during the secret recovery
phase can be protected by paired keys, so that the classic information will not
be leaked, and external attacks can be prevented.}

The organization of this paper is as follows: In Sect.2, we give
the preliminary knowledge related to bivariate polynomials and mutually unbiased
bases as well as the construction of unitary transformation. In Sect.3, we
proposed a new verifiable $(t,n)$-threshold QSS scheme with
sequential communication. The security of the scheme is analyzed in Section 4. Next,
we give a comparison of the basic properties among our scheme and others in Section 5.
Finally, the conclusion is given in Sect.6.

\section{Preliminaries}
\label{sec:2}

In this section, we  briefly introduce the knowledge related to bivariate
polynomials and mutually unbiased bases as well as the construction of unitary
transformation.

\subsection{Binary polynomials}
\label{subsec:2.1}

{\raggedright
A binary polynomial having degree $t - 1$ for $x$ and $h - 1$ for $y$ is defined as}

\begin{equation}
\begin{array}{l}
F\left( {x,y} \right) = {a_{00}} + {a_{10}}x + {a_{01}}y + {a_{11}}xy +{a_{20}}{x^2} +
 a{}_{02}{y^2} +{a_{12}}x{y^2} + {a_{21}}{x^2}y + {a_{22}}{x^2}{y^2} +\cdots  + \\
 {a_{t - 1,h -
1}}{x^{t - 1}}{y^{h - 1}}\bmod d,
\end{array}
\label{eq:1}
\end{equation}

{\raggedright
where ${a_{ij}} \in {{\cal F}_d},i \in \left\{ {0,1, \cdots t - 1} \right\},j \in
\left\{ {0,1, \cdots h - 1} \right\}$, ${x_i}$ is the public
information of ${\rm{Bo}}{{\rm{b}}_i}$. $d$ is required to be an odd
prime number in this paper. The verifiable secret sharing based on
binary polynomial is called BVSS. One advantage of BVSSs is that it can provide
shared keys for any two
participants ${\rm{Bo}}{{\rm{b}}_i}$ and ${\rm{Bo}}{{\rm{b}}_j}$ in the information
exchange process, so they can be protected. It can be divided into two
categories:
\par{(a).Verifiable secret sharing scheme based on symmetric bivariate
polynomial (SBVSSs)[15-18]. For SBVSSs, the dealer Alice chooses a symmetric
binary polynomial $F(x,y)$ and computes $F({x_i},y)\bmod d$ then sending it
to ${\rm{Bo}}{{\rm{b}}_i}$ through the secure channel. Any two
participants ${\rm{Bo}}{{\rm{b}}_i}$ and ${\rm{Bo}}{{\rm{b}}_j}$ compute
$F({x_i},{x_j})$ and $F({x_j},{x_i})$ respectively,using the pair as a shared key
between them.}
\par{(b).Verifiable secret sharing scheme based on asymmetric
bivariate polynomial (ABVSSs) [14,15]. For ABVSSs,
the dealer Alice chooses an asymmetric binary polynomial,
computing $F({x_i},y)\bmod d$,$F(x,{x_i})\bmod d$ and sending them
to ${\rm{Bo}}{{\rm{b}}_i}$ through the
secure channel. According to $F({x_i},y)\bmod d$ and
$F(x,{x_j})\bmod d$, ${\rm{Bo}}{{\rm{b}}_i}$ and ${\rm{Bo}}{{\rm{b}}_j}$ calculate
$F({x_i},{x_j})$ respectively as a pairwise shared key between them,
where $i,j \in \left\{ {1,2, \cdots ,n} \right\}$.}

\subsection{Mutually unbiased bases}
\label{subsec:2.2}

{\raggedright
Two sets of standard orthogonal bases ${A_1} = \left\{ {\left| {{\varphi _1}}
\right\rangle ,\left| {{\varphi _2}} \right\rangle , \cdots ,\left| {{\varphi
_d}} \right\rangle } \right\}$ and ${A_2} = \left\{ {\left| {{\psi _1}}
\right\rangle ,\left| {{\psi _2}} \right\rangle , \cdots ,\left| {{\psi _d}}
\right\rangle } \right\}$ are defined over a $d$-dimensional complex space  in Ref.
[19, 20] if the following relationship is satisfied:
}

\begin{equation}
\left| {\left\langle {{\varphi _i}} \right.\left| {{\psi _i}} \right\rangle }
\right| = \frac{1}{{\sqrt d }}.
\label{eq:2}
\end{equation}

{\raggedright
If any two of the set of standard orthogonal bases$\left\{ {{A_1},{A_2}, \cdots
,{A_m}} \right\}$ in space are unbiased, then this set is called an unbiased bases
set. Besides, it can be found $d + 1$ mutually unbiased bases if d is an odd prime
number.
}
First, the computation base is expressed as $\left\{ {\left| k \right\rangle
\left| {k \in D} \right.} \right\},D = \left\{ {0,1, \cdots ,d - 1} \right\}$,
and the remaining groups can be expressed as:

\begin{equation}
\left| {v_l^{\left( j \right)}} \right\rangle  = \frac{1}{{\sqrt d
}}\sum\limits_{k = 0}^{d - 1} {{w^{k\left( {l + jk} \right)}}\left| k
\right\rangle },
\label{eq:3}
\end{equation}

where $j$ and $l$ represent respectively the number of the mutually unbiased bases and the number of the vectors, $w = {e^{\frac{{2\pi i}}{d}}}$,$l,j \in D${\scriptsize . } These mutually unbiased bases satisfy the following conditions:

\begin{equation}
\left| {\left\langle {v_l^{\left( j \right)}} \right.\left| {v_l^{\left( {j'}
\right)}} \right\rangle } \right| = \frac{1}{{\sqrt d }},j \ne j'.
\label{eq:4}
\end{equation}

\subsection{The construction of unitary transformation}
\label{subsec:2.3}

Next, we introduce the two unitary transformations ${X_d}$ and ${Y_d}$ that we
need to use in this paper. In Ref.[13], they can be expressed as:

\begin{equation}
{X_d} = \sum\limits_{m = 0}^{d - 1} {{w^m}\left| m \right\rangle \left\langle m
\right|}.
\label{eq:5}
\end{equation}

Implementing \label{eq:5)} on $\left| {v_l^{\left( j \right)}} \right\rangle $ in turn, we
can obtain:

\begin{equation}
\begin{array}{l}
X_d^xY_d^y\left| {v_l^{\left( j \right)}} \right\rangle  = X_d^x\left(
{\sum\limits_{m = 0}^{d - 1} {{w^{y{m^2}}}\left| m \right\rangle \left\langle m
\right|} } \right)\left( {\frac{1}{{\sqrt d }}\sum\limits_{k = 0}^{d - 1}
{{w^{k\left( {l + jk} \right)}}\left| k \right\rangle } } \right)\\
\;\;\;\;\;\;\;\;\;\;\;\;\;\;\;\;\;\;\;\;= \frac{1}{{\sqrt d }}\sum\limits_{m = 0}^{d - 1}
{{w^{xm}}\left| m \right\rangle \left\langle m \right|} \sum\limits_{k = 0}^{d -
1} {{w^{k\left( {l + \left( {j + y} \right)k} \right)}}\left| k \right\rangle }
\\
\;\;\;\;\;\;\;\;\;\;\;\;\;\;\;\;\;\;\;\;= \frac{1}{{\sqrt d }}\sum\limits_{k = 0}^{d - 1}
{{w^{k\left( {\left( {l + x} \right) + \left( {j + y} \right)k} \right)}}\left| k
\right\rangle } \\
\;\;\;\;\;\;\;\;\;\;\;\;\;\;\;\;\;\;\;\;= \left| {v_{l + x}^{\left( {j + y} \right)}}.
\right\rangle
\end{array}
\label{eq:6}
\end{equation}

For the convenience of expression, $X_d^xY_d^y$ is denoted as ${U_{x,y}}$, that
is, ${U_{x,y}}\left| {v_l^{\left( j \right)}} \right\rangle  = \left| {v_{l +
x}^{\left( {j + y} \right)}} \right\rangle.$

\section{Verifiable $(t,n)$-threshold quantum secret sharing scheme}
\label{sec:3}
In this section, we construct a verifiable $(t,n)$-threshold quantum secret sharing scheme that includes an honest dealer Alice and shareholders ${\rm{Bo}}{{\rm{b}}_1}{\rm{,Bo}}{{\rm{b}}_2}, \cdots
,{\rm{Bo}}{{\rm{b}}_n}$.

\subsection{Preparation phase}
\label{subsec:3.1}

\paragraph{3.1.1} Alice randomly chooses an asymmetric bivariate polynomial of which the form is
like formula \label{eq:(1)} mentioned in the previous section 2, where ${a_{ij}} \in {{\cal
F}_d},i,j \in \left\{ {0,1, \cdots t - 1} \right\}$.
\paragraph{3.1.2} Alice calculates $F({x_i},y)$ and $F(x,{x_i})$ as the secret shares and sends it to
${\rm{Bo}}{{\rm{b}}_i}$ through secure channel, where ${x_i}$ is the public information
of ${\rm{Bo}}{{\rm{b}}_i}${\scriptsize .}
\paragraph{3.1.3} Alice prepares a $d$-dimensional quantum state $\left| \phi
\right\rangle  = \left| {\varphi _0^0} \right\rangle  = \frac{1}{{\sqrt d
}}\sum\limits_{i = 0}^{d - 1} {\left| i \right\rangle } $, and performs a unitary
transformation $U_{{p_0}},{{q_0}}$ on it, where ${p_0} = S$,${q_0} = s -
\sum\limits_{i = 1}^t {F\left( {{x_i},0} \right)} $,$s = F\left( {0,0}
\right)$,${p_0},{q_0} \in {{\cal F}_d}$, $S$ is a secret.

\subsection{Distribution phase}
\label{subsec:3.2}

\paragraph{3.2.1} Alice sends the quantum state ${\left| \phi  \right\rangle _0} =
{U_{{p_0},{q_0}}}\left| {\varphi _0^0} \right\rangle  = \left| {\varphi
_{{p_0}}^{{q_0}}} \right\rangle $ that she has performed a unitary transformation
on to ${\rm{Bo}}{{\rm{b}}_1}$ through the secure channel. Then
${\rm{Bo}}{{\rm{b}}_1}${\scriptsize  } performs a unitary
transformation ${U_{{p_1},{q_1}}}$ on the obtained quantum state ${\left| \phi
\right\rangle _0}$ to get ${\left| \phi  \right\rangle _1} =
{U_{{p_1},{q_1}}}\left| {\varphi _{{p_0}}^{{q_0}}} \right\rangle  = \left|
{\varphi _{{p_0} + {q_1}}^{{q_0} + {q_1}}} \right\rangle $, where ${p_1} = F\left(
{{x_1},{x_1}} \right)$,${q_1} = F\left( {{x_1},0} \right)$, ${p_1},{q_1} \in {{\cal
F}_d}$. Next, the quantum state ${\left| \phi  \right\rangle _1}$ performed
by ${\rm{Bo}}{{\rm{b}}_1}$ is sent to ${\rm{Bo}}{{\rm{b}}_2}$.
\paragraph{3.2.2} The other participants ${\rm{Bo}}{{\rm{b}}_i}${\scriptsize  } repeat the same
operation of ${\rm{Bo}}{{\rm{b}}_1}${\scriptsize  } in 3.2.1 , that is
, ${\rm{Bo}}{{\rm{b}}_i}${\scriptsize  } performs a unitary
transformation ${U_{{p_i},{q_i}}}$ on the obtained quantum state ${\left| \phi
\right\rangle _{i - 1}}$ from ${\rm{Bo}}{{\rm{b}}_{i - 1}}$, getting

\begin{equation}
{\left| \phi  \right\rangle _i} = {U_{{p_i},{q_i}}}\left| {\varphi
_{\sum\limits_{i = 0}^{i - 1} {{p_i}} }^{\sum\limits_{i = 0}^{i - 1} {{q_i}} }}
\right\rangle  = \left| {\varphi _{\sum\limits_{i = 0}^i {{p_i}}
}^{\sum\limits_{i = 0}^i {{q_i}} }} \right\rangle.
\label{eq:7}
\end{equation}

Then he sends the quantum state ${\left| \phi
\right\rangle _{i }}$ to ${\rm{Bo}}{{\rm{b}}_{i + 1}}$ through the
quantum safe channel until the last participant ${\rm{Bo}}{{\rm{b}}_t}$ performs
the same operation and gets the final state

\begin{equation}
{\left| \phi  \right\rangle _t} = {U_{{p_t},{q_t}}}\left| {\varphi
_{\sum\limits_{i = 0}^{t - 1} {{p_i}} }^{\sum\limits_{i = 0}^{t - 1} {{q_i}} }}
\right\rangle  = \left| {\varphi _{\sum\limits_{i = 0}^t {{p_i}}
}^{\sum\limits_{i = 0}^t {{q_i}} }} \right\rangle,
\label{eq:8}
\end{equation}

where ${p_i} = F({x_i},{x_i})$,${q_i} = F({x_i},0),{p_i},{q_i} \in {{\cal F}_d}$,$i =
2,3, \cdots ,t${\scriptsize .}

\paragraph{3.2.3} Since ${k_{ij}} = F({x_i},{x_j})$ is used as a shared key between participants ${\rm{Bo}}{{\rm{b}}_i}$ and ${\rm{Bo}}{{\rm{b}}_j}$, ${\rm{Bo}}{{\rm{b}}_i}$ can calculate ${c'_i}  = {E_{{k_{ij}}}}({p_i})$,${c''_i} ={E_{{k_{ij}}}}({q_i})$ and send them to ${\rm{Bo}}{{\rm{b}}_j}$. After
 receiving the ciphertext ${c'_i}$,${c''_i}$, he can infer ${p_i} = {D_{{k_{ij}}}}({c'_i})$,${q_i} = {D_{{k_{ij}}}}({c''_i})$, where
${E_{{k_{ij}}}}({p_i})$,${E_{{k_{ij}}}}({q_i})$ represents classical encryption of
plaintext ${p_i},{q_i}$, ${D_{{k_{ij}}}}({c'_i})$,${D_{{k_{ij}}}}({c''_i})$ represents classical
decryption of ciphertext ${c'_i},{c''_i}$, $i,j \in \left\{ {1,2, \cdots ,t}
\right\}$,$i \ne j$.

\subsection{Measurement phase}
\label{subsec:3.3}

\paragraph{3.3.1} After receiving ${p_i},{q_i}$, by the binary Lagrange interpolation formula, ${\rm{Bo}}{{\rm{b}}_j}$ can calculate

\begin{equation}
s' = \sum\limits_{i = 1}^t {\left( {F\left( {{x_i},0}
\right)\prod\limits_{\scriptstyle k = 1\hfill\atop
\scriptstyle k \ne i\hfill}^t {\frac{{{x_k}}}{{{x_k} - {x_i}}}} } \right)}  =
\sum\limits_{i = 1}^t {\left( {{q_i}\prod\limits_{\scriptstyle k = 1\hfill\atop
\scriptstyle k \ne i\hfill}^t {\frac{{{x_k}}}{{{x_k} - {x_i}}}} } \right)}
\label{eq:9}
\end{equation}

where $i,j \in \left\{ {1,2, \cdots ,t} \right\}${\scriptsize .} At first, the
last participant ${\rm{Bo}}{{\rm{b}}_t}$ performs a unitary transformation
${U_{{p_t},{q_t}}}$ on the obtained quantum state ${\left| \phi  \right\rangle
_{t - 1}}$ to obtain ${\left| \phi  \right\rangle _t}$, and then he select bases
$\left\{ {\left| {v_l^{\left( {s'} \right)}} \right\rangle } \right\}$ to measure
the quantum state ${\left| \phi  \right\rangle _t}$ with the measurement result
denoted as $R'${\scriptsize . } Next he could calculate $c = {E_{{k_{ti}}}}(R')$
and sent it to other participants ${\rm{Bo}}{{\rm{b}}_i}$.
\paragraph{3.3.2} After receiving the ciphertext $c${\scriptsize  } from the last
participant ${\rm{Bo}}{{\rm{b}}_t}$, ${\rm{Bo}}{{\rm{b}}_i}${\scriptsize
} calculates $R' = {D_{{k_{ti}}}}(c)$, where ${E_{{k_{ti}}}}(R')$ is the classic
encryption of $R'$ sent by ${\rm{Bo}}{{\rm{b}}_t}$, and ${D_{{k_{ti}}}}(c)$ is the
classic decryption of ciphertext $c$, where $i = 1,2, \cdots ,t - 1${\scriptsize
.}

\subsection{Testing phase}
\label{subsec:3.4}

\paragraph{3.4.1} For the security of the scheme, the quantum states are randomly selected and
detected by Alice during the process of transmitting the quantum states. Alice
requires ${\rm{Bo}}{{\rm{b}}_i}${\scriptsize  } to send the calculated $s'$ to her
and checks whether it is satisfied. If satisfied, the participants are honest and
the scheme continues because the following formula is true:

\begin{equation}
\sum\limits_{j = 0}^t {{q_i} = \left( {s - \sum\limits_{j = 1}^t {{q_i}} }
\right) + } \sum\limits_{j = 1}^t {{q_j}} \bmod d = s.
\label{eq:10}
\end{equation}

If one or some of the participants calculate $s' = s$ is not satisfied, it
indicates that the participant has fraudulent behavior, which can be divided into
two cases. If $s' = s$ calculated by the last participant is not satisfied, the
scheme terminates. If it is found that one or some of the previous
participants ${\rm{Bo}}{{\rm{b}}_1}{\rm{,Bo}}{{\rm{b}}_2}{\rm{,}} \cdots
{\rm{,Bo}}{{\rm{b}}_{t - 1}}$ dissatisfy $s' = s$, then it move on to the next
step.
\paragraph{3.4.2} As long as Alice checks each two participant's ${p_i},{p_j}$ that send
by ${\rm{Bo}}{{\rm{b}}_i}$ and received by ${\rm{Bo}}{{\rm{b}}_j}$, she can find out
which participant has fraudulent behavior, and remove it in the next round of
secret sharing scheme, where $i,j \in \left\{ {1,2, \cdots ,t - 1} \right\},i \ne
j$.
\paragraph{3.4.3} Alice checks each participant's received $R'$ sent by the last participant
${\rm{Bo}}{{\rm{b}}_t}$ and examine if the following is established:

\begin{equation}
R' = R = {p_0} + {p_1} +  \cdots  + {p_t} = S + F\left( {{x_1},{x_1}} \right) +
\cdots  + F\left( {{x_t},{x_t}} \right).
\label{eq:11}
\end{equation}

If it is established, the scheme continues; if at least one participant
receives $R'$ such that \label{eq:9)} does not hold, indicating that the last participant
${\rm{Bo}}{{\rm{b}}_t}$ is dishonest and the scheme is terminated,
so ${\rm{Bo}}{{\rm{b}}_t}$ is removed in the next round of secret sharing.

\subsection{Recovery phase}
\label{subsec:3.5}

In order to restore the original secret, ${\rm{Bo}}{{\rm{b}}_i}$ can calculate

\begin{equation}
{p_0} = R - \sum\limits_{i = 1}^t {{p_i}}
\label{eq:12}
\end{equation}

and obtain the secret $p_0=S$. If Alice wants to share multiple secrets, then repeat the
above process, where $i = 1,2, \cdots ,t$.

\section{Correctness and security}
\label{sec:4}

In this section, we mainly account for the correctness analysis and the security
of our scheme against four primary attack: dishonest
participant attacks ,the intercept-and-resend attack,
entangle-and-measure attack and collusion attack.

\subsection{Correctness analysis}
\label{subsec:4.1}

After all participants ${\rm{Bo}}{{\rm{b}}_1},{\rm{Bo}}{{\rm{b}}_2}, \cdots
,{\rm{Bo}}{{\rm{b}}_t}$ complete their operations, the final quantum state is

\begin{equation}
{\left| \phi  \right\rangle _t} = \left( {\prod\limits_{k = 0}^t
{{U_{{p_k},{q_k}}}} } \right)\left| \phi  \right\rangle  = \left| {\varphi
_{\sum\limits_{k = 0}^t {{p_k}} }^{\sum\limits_{k = 0}^t {{q_k}} }} \right\rangle.
\label{eq:12}
\end{equation}

Based on the binary Lagrange interpolation formula, they can calculate $s$ after
exchanging classic information ${p_i},{q_i}$. The last participant
${\rm{Bo}}{{\rm{b}}_t}$ selects the basis to measure the final state and the
measured result $R = \sum\limits_{i = 0}^t {{p_i}} $ is sent to each participant
after being encrypted by the shared key, so each participant can recover the
secret ${p_0} = S$, where $s = \sum\limits_{i = 0}^t {{q_i}} \bmod d$.

\subsection{Security analysis}
\label{subsec:4.2}
The security of the scheme is analyzed in this section.

\subsubsection{Participant attack}
\label{subsubsec:4.2.1}

One or some of the participants use the random numbers replace the real ones for
the unitary operation during the unitary transformation phase. It is checked
whether $s'$ calculated by each participant satisfies $s'=s$ in the testing phase
is true, so that dishonest participants are found. This indicates that the participant
attack is invalid. Therefore the secret cannot be recovered.

\subsubsection{Intercept-and-resend attack}
\label{subsubsec:4.2.2}

We suppose that the eavesdropper Eve intercepts the quantum state $\left|
{\varphi _l^k} \right\rangle $ sent
by ${\rm{Bo}}{{\rm{b}}_i}$ to ${\rm{Bo}}{{\rm{b}}_{i + 1}}$, but he does not know
any information about the measurement basis. He can only choose the correct
measurement basis with the probability of ${1 \mathord{\left/
{\vphantom {1 d}} \right.
\kern-\nulldelimiterspace} d}$. In addition, the measurement result is

\begin{equation}
S + \sum\limits_{k = 1}^{i - 1} {{p_i}}.
\label{eq:13}
\end{equation}

If Eve does not know the basis chosen by the participant before, he can only
infer the secret of the dealer with the probability of ${1 \mathord{\left/
{\vphantom {1 d}} \right.
\kern-\nulldelimiterspace} d}$. In short, Eve cannot obtain the secret with a
probability of exceeding${1 \mathord{\left/
{\vphantom {1 d}} \right.
\kern-\nulldelimiterspace} d}$ in intercept-and-resend attack.

\subsubsection{Entangle-and-measure Attack}
\label{subsubsec:4.2.3}

The eavesdropper Eve entangles the auxiliary quantum state onto the transmitted
quantum state or replaces the quantum state with a new entangled state, but the
entanglement exchange causes the quantum state to be in a mixed state. There is
no way to distinguish, so he cannot obtain any information about the measurement
basis and participant's ${p_i},{q_i}$. This will also make formula \label{eq:10)} and
\label{eq:11)} are false, so it will be found by Alice.

\subsubsection{Collusion attack}
\label{subsubsec:4.2.4}
Participant collusion is a more destructive attack than an external attack. It
is assumed that in the worst case only the dealer Alice and one participant are
honest, and the remaining $t - 1$ participants will conduct a collusion attack
that they exchange their ${p_i}$ only to get $\sum\limits_{i = 1}^{t - 1} {{p_i}} $
but could not get the original secret ${p_0}$. Therefore, it is invalid with
collusion attack.

\section{Comparison}
\label{sec:5}

Here, we give a comparison of the basic properties in our scheme and other
$d$-dimensional QSS schemes.
The dealer in Lu's scheme [9] can share three secrets by delivering three
identical states sequentially among participants. To recover the secret, they
perform proper unitary operations on a vector of a set of MUBs and the qubits are
measured in an appointed basis by the last participant. After that, they exchange
the random numbers that are embedded in the qubits and vulnerable to eavesdropper
to recover the secrets. Besides, a verifiable $(t,n)$-threshold
quantum secret sharing scheme is proposed by using-dimensional Bell state and the
Lagrange interpolation. The scheme is verified by Hash function with less
verification information. We give a new verifiable QSS scheme based on single
qubit in this paper by using sequential communication of a single quantum
$d$-dimensional system and the Lagrange interpolation. A detailed
comparison of these schemes is presented in Table 1.

\begin{table}
\centering
\caption{comparison among QSS schemes}
\label{tab:1}       
\begin{tabular}{p{107pt}p{88pt}p{80pt}p{90pt}}
\hline
\parbox{107pt}{\raggedright \label{OLE_LINK1}} & \parbox{88pt}{\raggedright
{\tiny Lu[9]}
} & \parbox{80pt}{\raggedright
{\tiny Qin[20]}
} & \parbox{90pt}{\raggedright
{\tiny New}
} \\
\hline
\parbox{107pt}{\raggedright
{\tiny Access structure}
} & \parbox{88pt}{\raggedright
$(t,n)${\tiny -threshold}
} & \parbox{80pt}{\raggedright
{\tiny -threshold}
} & \parbox{90pt}{\raggedright
$(t,n)${\tiny -threshold}
} \\
\parbox{107pt}{\raggedright
{\tiny Dimension of space}
} & \parbox{88pt}{\raggedright $d$}
 & \parbox{80pt}{\raggedright $d$}
 & \parbox{90pt}{\raggedright $d$} \\
\parbox{107pt}{\raggedright
{\tiny Quantum state}
} & \parbox{88pt}{\raggedright
{\tiny Three qubits}
} & \parbox{80pt}{\raggedright
{\tiny Two qubits}
} & \parbox{90pt}{\raggedright
{\tiny Single qubit}
} \\
\parbox{107pt}{\raggedright
{\tiny Quantum operations}
} & \parbox{88pt}{\raggedright
{\tiny The unitary operation}
} & \parbox{80pt}{\raggedright
{\tiny The unitary operation}
} & \parbox{90pt}{\raggedright
{\tiny The unitary operation}
} \\
\parbox{107pt}{\raggedright
{\tiny Method}
} & \parbox{88pt}{\raggedright
{\tiny Lagrange interpolation, MUB}
} & \parbox{80pt}{\raggedright
{\tiny Lagrange interpolation}
} & \parbox{90pt}{\raggedright
{\tiny Lagrange interpolation, MUB}
} \\
\parbox{107pt}{\raggedright
{\tiny Verification of secret}
} & \parbox{88pt}{\raggedright
{\tiny Verifiable equation}
} & \parbox{80pt}{\raggedright
{\tiny Hash function}
} & \parbox{90pt}{\raggedright
{\tiny Supervision of the dealer}
} \\
\parbox{107pt}{\raggedright
{\tiny Protection of classic information}
} & \parbox{88pt}{\raggedright
{\tiny NO}
} & \parbox{80pt}{\raggedright
{\tiny -}
} & \parbox{90pt}{\raggedright
{\tiny Yes}
} \\
\hline
\end{tabular}
\end{table}

So it can be seen that
\par{(a) Our scheme achieves the verifiability with less verification information under
the supervision of the dealer Alice.}
\par{(b)The scheme adds shared keys to protect the share information that needs to be
exchanged, thereby making this scheme be more secure for external attacks.}

\section{Conclusion}
\label{sec:6}

In this paper, a new verifiable $(t,n)$-threshold quantum secret sharing scheme based on binary polynomial and mutually unbiased bases is proposed in combination with the Ref.[9]. Compared with the original scheme, our scheme makes a dealer supervise whether each participant is honest, thereby achieving verifiability. If the dishonest participants were found during the testing phase, he can be eliminated to prevent internal attacks. This paper also uses a binary polynomial to add a pair of shared keys to ensure confidentiality. Participants can be protected by paired keys when exchanging classic information in the recovery phase, which is more secure than the original one with preventing external attacks. The security analysis illuminates that our scheme can resist the participants attack, the intercept-and-resend attack, the entangle-and-measure attack and the collusion attack. Therefore, the security of the scheme is significantly improved. Compared with other existing verifiable schemes, the verifiable mechanism based on the supervision of the dealer is implemented in this paper, which will not waste extra quantum states with less authentication information.

\paragraph{Acknowledgements}
We would like to thank anonymous reviewer for valuable comments. This work is
supported by the National Natural Science Foundation of China
under Grant No.11671244.

\end{document}